\documentclass[journal]{IEEEtran}
\usepackage{cite}

\newcommand{\instbit}[1]{\mbox{\scriptsize #1}}

\ifCLASSINFOpdf
  \usepackage[pdftex]{graphicx}
   \graphicspath{{./figures/}}
\else
  \usepackage[dvips]{graphicx}
   \graphicspath{{./figures/}}
\fi
\usepackage{amsmath}
\usepackage{amsfonts}
\usepackage{array}
\usepackage{makecell}
\usepackage{multicol}
\usepackage{multirow}
\usepackage{threeparttable}

\usepackage{float}
\usepackage{stfloats}
\usepackage{url}

\usepackage{hyperref}

\usepackage{pifont}
\usepackage{upgreek}
\usepackage{xcolor}

\begin{document}
%
\title{A Customized NoC Architecture to Enable Highly Localized Computing-On-the-Move DNN Dataflow}
%
%
%

\author{Kaining~Zhou,~\IEEEmembership{Student~Member,~IEEE},
        Yangshuo~He,~\IEEEmembership{Student~Member,~IEEE},
        Rui~Xiao,~\IEEEmembership{Student~Member,~IEEE},
        Jiayi~Liu,~\IEEEmembership{Student~Member,~IEEE},
        Kejie~Huang,~\IEEEmembership{Senior~Member,~IEEE}
\thanks{This work was supported by the Major Scientific Research Project of Zhejiang Lab (No. 2019KC0AD02), the National Natural Science Foundation of China (U19B2043), and the National Key Research and Development Program of China (2020AAA0109002). \textit{(Kaining~Zhou and Yangshuo~He are co-first authors.)} \textit{(Corresponding author: Kejie~Huang.)}

Authors are with the College of Information Science \& Electronic Engineering, Zhejiang University, 38 Zheda Road, Hangzhou, China, 310027, K. Huang is also with Zhejiang Lab, Building 10, China Artificial Intelligence Town, 1818 Wenyi West Road, Hangzhou City, Zhejiang Province, China, email: zkn.gml@gmail.com; sugarhe@zju.edu.cn; xiaor@zju.edu.cn; 3170105617@zju.edu.cn, huangkejie@zju.edu.cn

}
}

%
%

\markboth{IEEE TRANSACTIONS ON CIRCUITS AND SYSTEMS II: EXPRESS BRIEFS}%
{Shell \MakeLowercase{\textit{et al.}}: Bare Demo of IEEEtran.cls for IEEE Journals}
%



\maketitle

\begin{abstract}
The ever-increasing computation complexity of fast-growing Deep Neural Networks (DNNs) has requested new computing paradigms to overcome the memory wall in conventional Von Neumann computing architectures. The emerging Computing-In-Memory (CIM) architecture has been a promising candidate to accelerate neural network computing. However, data movement between CIM arrays may still dominate the total power consumption in conventional designs. This paper proposes a flexible CIM processor architecture named Domino and ``Computing-On-the-Move'' (COM) dataflow, to enable stream computing and local data access to significantly reduce data movement energy. Meanwhile, Domino employs customized distributed instruction scheduling within Network-on-Chip (NoC) to implement inter-memory computing and attain mapping flexibility. The evaluation with prevailing DNN models shows that Domino achieves 1.77-to-2.37$\times$ power efficiency over several state-of-the-art CIM accelerators and improves the throughput by 1.28-to-13.16$\times$.
\end{abstract}


%
\IEEEpeerreviewmaketitle

\section{Introduction}
%
%
%
%

\IEEEPARstart{T}{he} rapid development of Deep Neural Network (DNN) algorithms has led to high energy consumption due to millions of parameters and billions of operations in one inference\cite{senet}\cite{vggnet}. Conventional processors such as CPUs and GPUs are power-hungry devices and inefficient for AI computations. Therefore, accelerators that improve computing efficiency are under intensive development to meet the power requirement in the post-Moore's Law era.

One of the most promising solutions is to adopt Computing-In-Memory (CIM) scheme to increase the parallel computation speed with much lower computation power. Recently, both volatile memory and non-volatile memory have been proposed as computing memories for CIM\cite{ISSCC_2021_Yoon}\cite{isscc-digital-pim}. However, existing works mainly focus on the design of CIM arrays but lack a flexible top-level architecture for configuring storage and computing units of DNNs. These designs need to access off-chip memory frequently, leading to high power consumption and long latency. Therefore, new flexible top-level architecture and efficient dataflow should be studied to meet various requirements of DNNs while achieving high hardware resource utilization and energy efficiency.

Network-on-Chip (NoC) with high parallelism and scalability has attracted lots of attention. In particular, NoC can optimize the process of computing DNN algorithms by organizing multiple cores uniformly under specified hardware architectures\cite{Eyerissv2}\cite{eyerissv1}. The conventional NoC based CIM architectures such as \cite{ISSCC_2021_Jia} are inefficient for various convolution kernel sizes and need to load the input activation multiple times. This paper proposes a customized NoC architecture called Domino to enable highly localized inter-memory computing for DNN inference to minimize data reload. Inter-memory computing means that computing like partial sum addition, activation, and pooling is performed in the network when data are moving between CIM arrays. Consequently, ``Computing-On-the-Move'' (COM) dataflow is proposed to maximize data locality and significantly reduce the energy of data movement. Dataflow is controlled by distributed local instructions instead of an external/global controller or processor. Evaluation results show Domino improves power efficiency and throughput by more than 77\% and 28\%, respectively.

The rest of the paper is organized as follows: \autoref{sec:architecture} describes the architecture and building blocks of Domino; \autoref{sec:dataflow} illustrates the dataflow model; \autoref{sec:evaluation} presents the evaluation setup, experimental results and comparisons; finally, \autoref{conclusion} draws the conclusion.

\section{Domino Architecture}
\label{sec:architecture}

From a top view, Domino mainly consists of an array of tiles interconnected in a 2-D mesh NoC. Weights of each layer (e.g., convolution (CONV) and fully connected (FC) layer) of a neural network will be mapped to a certain group of tiles on Domino, as shown in \autoref{fig:tile} (a). By this means, Domino achieves a flexible and distributed computation architecture for DNN acceleration.

\subsection{Domino Tile}

\begin{figure}[ht]
    \centering  
    \includegraphics[width=0.45\textwidth]{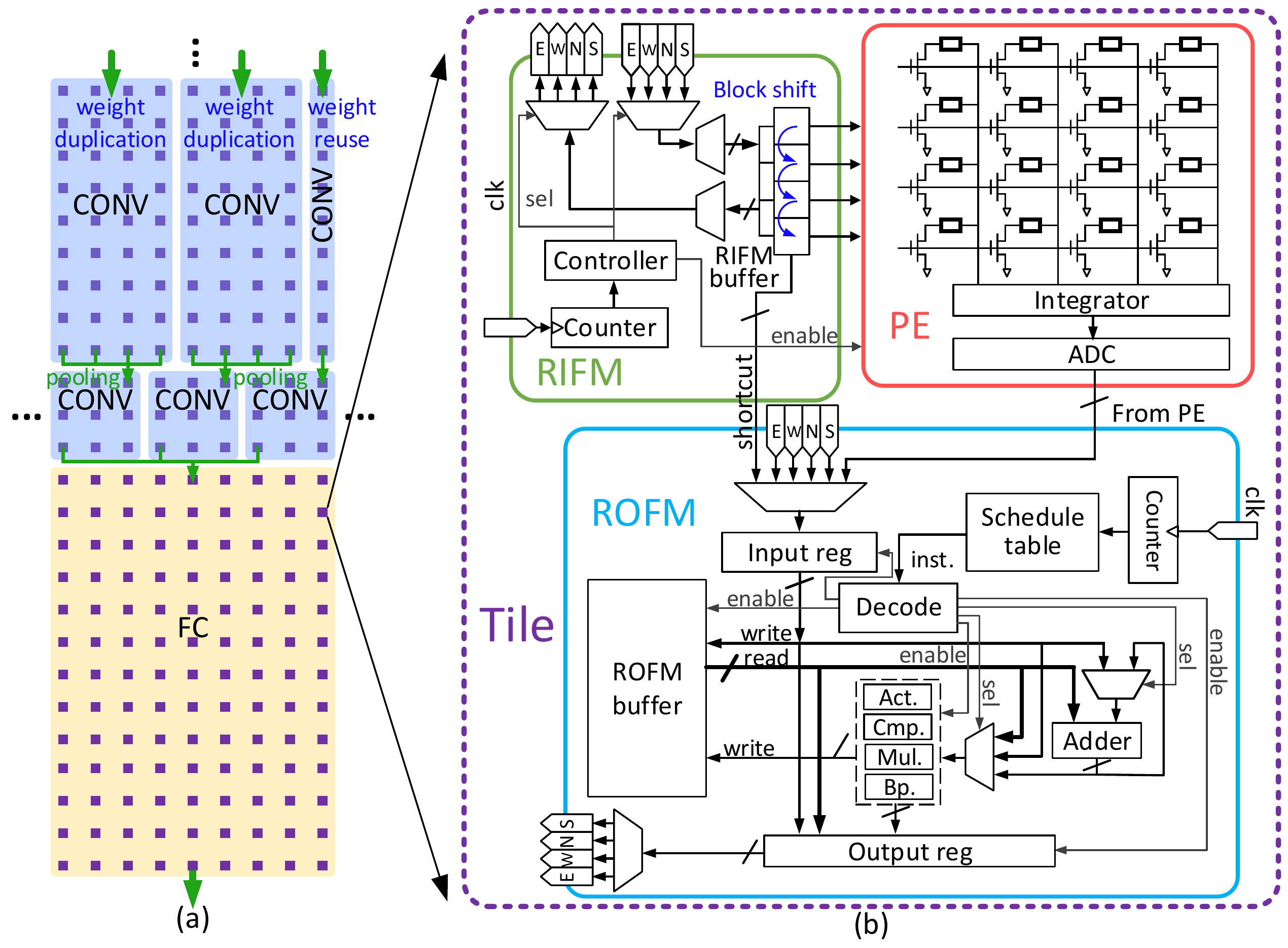}
    \caption{(a) Each tile array is mapped with a layer in the neural network; (b) A Domino tile contains an RIFM, an ROFM and a computation center PE.}  
    \label{fig:tile}
\end{figure}

A tile includes a CIM array called a Processing Element (PE), a router transferring Input Feature Maps (IFMs) called an RIFM, and a router transferring Output Feature Maps (OFMs) and partial-sums in convolution computation called an ROFM. The basic structure of a tile is illustrated in \autoref{fig:tile} (b). The RIFM receives input data from one out of four directions in each tile and controls input dataflow to a remote RIFM, the local PE, and the local ROFM. In-memory computing usually starts from the RIFM buffer and ends at Analog-to-Digital Converters (ADC) in a PE (if adopting ReRAM-based CIM arrays). Outputs of a PE are sent to an ROFM for temporary storage or partial-sum addition. The ROFM is controlled by periodic instructions to receive either computation results or input data via a shortcut from an RIFM and maintain dataflow to add up partial-sums.

\subsection{Domino RIFM}
\label{sec:RIFM}

As shown in \autoref{fig:tile} (b), each RIFM possesses I/O ports in four directions to communicate with RIFMs in the adjacent tiles. It is also equipped with a buffer called RIFM buffer to store received input data in the current cycle. Data in RIFM buffers are sent to the PE for Multiplication-and-Accumulation (MAC) computations. It supports an in-buffer shifting operation with a step size of $64$ or a multiple of $128$. The in-buffer shifting architecture maximizes in-tile data reuse when handling the first few layers with small input channel numbers. A shortcut connection from the RIFM to the ROFM is established to support the situation that MAC computation is skipped (i.e., the shortcut in a ResUnit). A counter and a controller in the RIFM decide input dataflow based on the initial configuration. Once the RIFM receives input packets, the counter starts to increase its value. With COM dataflow, no input matrix conversion like im2col\cite{im2col} is required. Details of COM dataflow will be introduced in Section \ref{sec:dataflow}. 

\subsection{Domino ROFM}
The ROFM is the key component for COM dataflow controlled by instructions to manage I/O ports and buffers, add up partial/group-sum results, and perform activation or pooling to get convolution results. \autoref{fig:tile} (b) shows its micro-architecture consisting of a set of four-direction I/O ports, input/output registers, an instruction schedule table, a counter to generate instruction indices, an ROFM buffer to store partial computation results, reusable adders, a computation unit with adequate functions, and a decoder. 

The ROFM is configured and ruled by localized instructions fitting inter-memory dataflow to support the COM procedure. The compiler generates instructions and configuration for each tile based on initial input data and the DNN structure. The instruction format is shown in \autoref{tab:instruction}. Details about inter-memory computing functions are listed in \autoref{tab:ROFM_function}.

\begin{table}[h]
    \caption{The instruction format for Domino.}
    \begin{small}
        \begin{center}
            \begin{tabular}{p{0in}p{0.05in}p{0.05in}p{0.05in}p{0.05in}p{0.05in}p{0.05in}p{0.05in}p{0.05in}l}
            \multicolumn{1}{l}{\instbit{15}} &
            \multicolumn{1}{r}{\instbit{11}} &
            
            \multicolumn{1}{l}{\instbit{10}} &
            \multicolumn{1}{l}{\instbit{7}} &
            
            \multicolumn{1}{l}{\instbit{6}} &
            \multicolumn{1}{l}{\instbit{5}} &
            
            \multicolumn{1}{l}{\instbit{4}} &
            \multicolumn{1}{l}{\instbit{1}} &
            \multicolumn{1}{r}{\instbit{0}} &  \\
            
            \cline{1-9}
            
            \multicolumn{2}{|c|}{Rx Ctrl.} &
            \multicolumn{2}{c|}{Sum} &
            \multicolumn{2}{c|}{Buffer} &
            \multicolumn{2}{c|}{Tx Ctrl.} &
            \multicolumn{1}{c|}{Opc.} &  C-type \\
            
            \cline{1-9}
            
            \multicolumn{2}{|c|}{Rx Ctrl.} &
            \multicolumn{4}{c|}{Func.} &
            \multicolumn{2}{c|}{Tx Ctrl.} &
            \multicolumn{1}{c|}{Opc.} &  M-type \\
            
            \cline{1-9}
            
            \end{tabular}
        \end{center}
    \end{small}
    \label{tab:instruction}
\end{table}

\begin{table}[hbp]
    \caption{\textcolor{black}{Inter-memory computing functions supported by ROFMs.}}
    \begin{center}
        \scriptsize
            \begin{tabular}{|c|c|c|}
            \hline
                Function & Explanation                        & Usage                    \\
                \hline
                Add      & Adder                              & Partial sum accumulation \\
                \hline
                Act.     & Activation                         & Non-linear function      \\
                \hline
                Cmp.     & Comparison                         & Max pooling              \\
                \hline
                Mul.     & Multiplication with a scaling factor & Average pooling          \\
                \hline
                Bp.      & Direct transmission                & ``Skip'' connection      \\
                \hline
            \end{tabular}
        \label{tab:ROFM_function}
    \end{center}
\end{table}

After cycle-accurate analyses and mathematical derivation, instructions reveal an attribute of periodicity. During the convolution computation, C-type instructions are fetched from the schedule table and executed periodically. When convolution stride $S_c = 1$, the period $p = 2(P+W)$ ($P$ is the padding size and $W$ is the width of the IFM) is determined by neural network configuration. When $S_c \neq 1$, the compiler will shield certain bits in control words to ``skip'' some actions in the corresponding cycles to make a correct computation. When an ROFM is mapped and configured to process the last row of a layer in a Convolution Neural Network (CNN), it will generate activation and pooling instructions of M-type. Its period is related to pooling stride $S_p$ ($p=2S_p$). The instructions for pooling layers and FC layers are also periodic.

Partial-sums are added to group-sums when they are transferred between tiles. The group-sums are queued in the buffer for other group-sums to be ready and then form a complete computation result. This method enables inter-memory computing when data are moving between tiles. With localized and customized instructions, Domino manages to reduce the bandwidth demand for transmitting data or instructions through NoC, while maintaining the flexibility for various DNNs.

\subsection{Domino PE}
\label{sec:PE}

Our main focus is top-level architecture and dataflow rather than the design of CIM cores. Therefore, Domino adopts existing CIM arrays to enable flexible substitution. In our design, each crossbar array has $N_c$ rows and $N_m$ columns.

\section{Dataflow Model}
\label{sec:dataflow}
Weight Stationary (WS), Output Stationary (OS), and Row Stationary (RS) are three widely used dataflows\cite{Efficient}. However, conventional dataflows are inefficient for CIM schemes. In this paper, we propose COM dataflow based on WS dataflow to reduce data movement for both partial-sums and IFMs. COM dataflow is customized for CIM architecture with two innovative features: (1) though weights are stationary, the conversion from IFMs to Toeplitz matrices is not required for convolution. Similar to RS dataflow, input activations are reused among different tiles. Therefore, there is no data reload or duplication and the data movement of IFMs is minimized. While in \cite{ISSCC_2021_Jia}, IFMs and weights must be loaded repeatedly during runtime. (2) Partial- and group-sums are stored in tile buffers instead of external global buffers, greatly reducing energy for data movement. Partial-sums are accumulated in the local buffer or when they are transmitting along the array of routers, further minimizing data movement of partial-sums.

\subsection{Dataflow in FC layers}
\label{sec:fc}

\begin{figure}[htb]
    \centering  
    \includegraphics[width=0.45\textwidth]{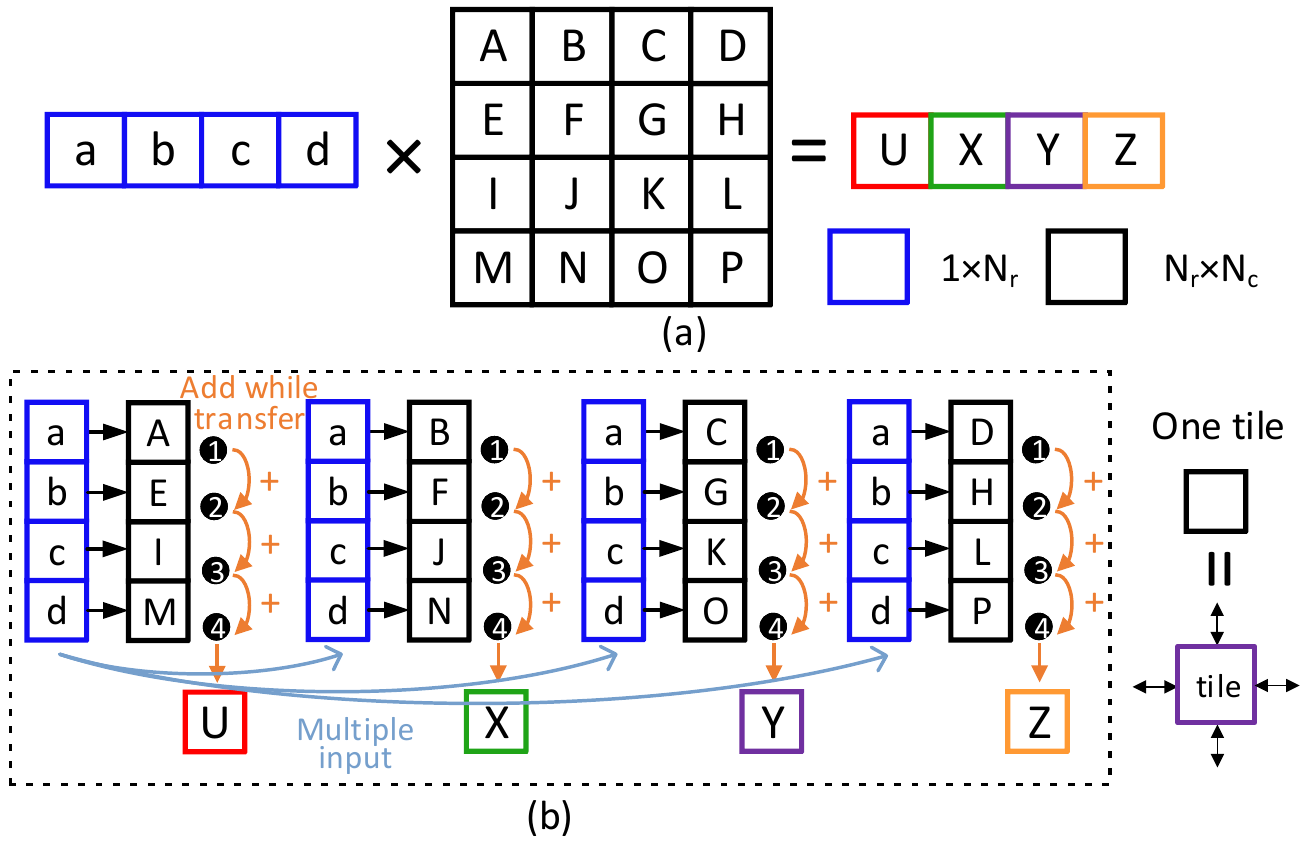}
    \caption{The proposed mapping and dataflow in FC layers: (a) the partitioned input vector and weight matrix; (b) the dataflow to transmit and add partial sums to a complete  result.}  
    \label{fig:fc_dataflow}
\end{figure}

FC layers perform Matrix-Vector Multiplication (MVM) which can be formulated as $\mathbf{y} = \mathbf{xW}$, where $\mathbf{x}\in  \mathbb{R}^{1\times C_{in}}$, $\mathbf{y}\in  \mathbb{R}^{1\times C_{out}}$, and $\mathbf{W}\in \mathbb{R}^{C_{in}\times C_{out}}$ are input vector, output vector, and weight matrix, respectively. In most cases, an $N_c\times N_m$ crossbar array is insufficient to map the complete weight matrix in an FC layer. Therefore, an array of tiles with $\lceil\frac{C_{in}}{N_c}\rceil$ rows and $\lceil\frac{C_{out}}{N_m}\rceil$ columns is allocated to efficiently handle Blocked Matrix Multiplication (BMM).

COM dataflow in FC layers is similar to WS dataflow in systolic arrays, but without weights reload during computation. As shown in \autoref{fig:fc_dataflow}, each tile maps to a block matrix and the PE multiplies weight with a slice of input. Multiplication results (\ding{182} to \ding{185}) are added while transmitting along a column of tiles. Final addition results in the last tiles of four columns, $\mathbf{U}$ to $\mathbf{Z}$, are small slices of an output vector. Concatenating small slices in all columns gives the complete BMM result.

\begin{figure}[htb]
    \centering  
    \includegraphics[width=0.47\textwidth]{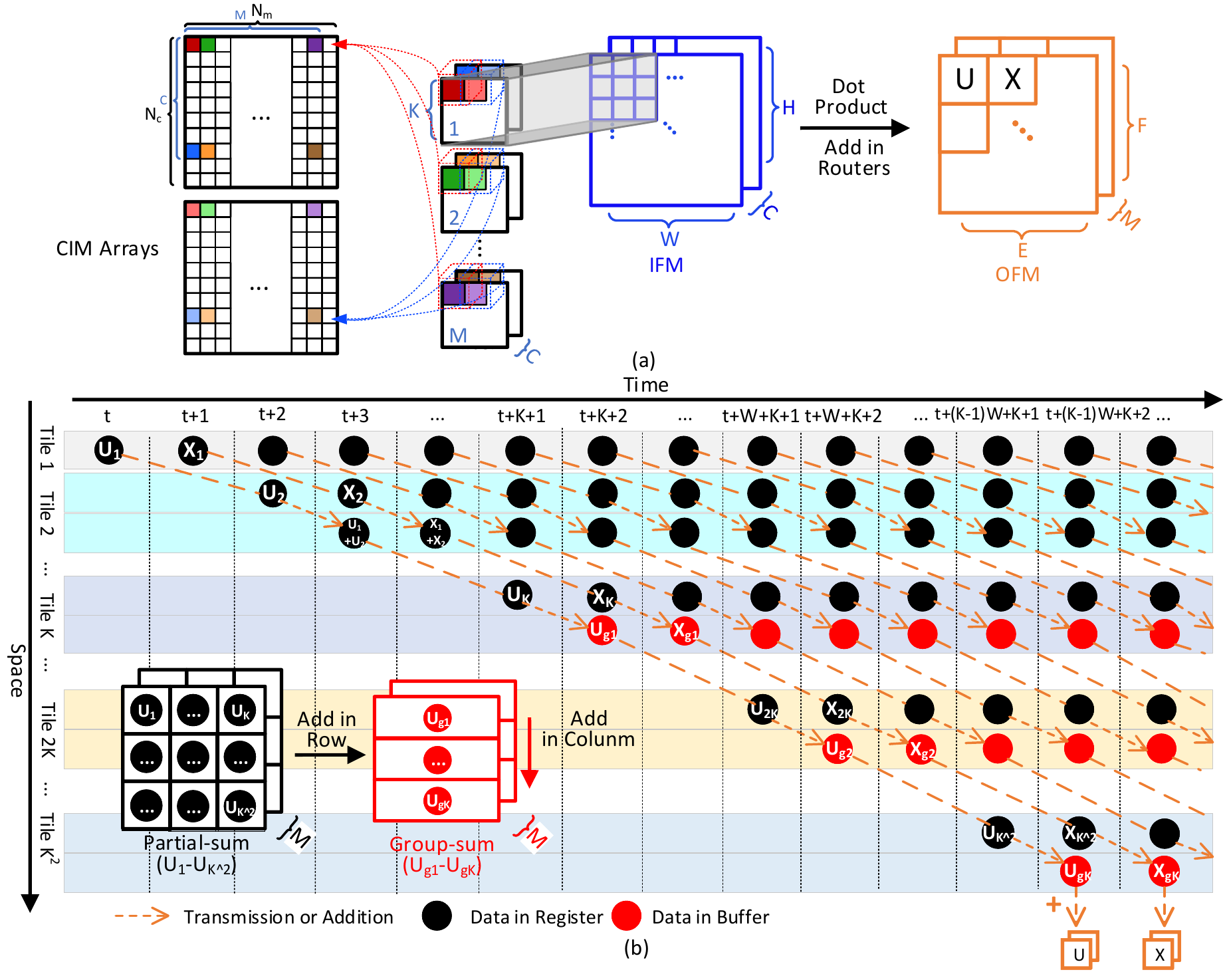}
    \caption{(a) The mapping strategy and the computation process. (b) The timing and location of COM dataflow in CONV layer, black circles represent partial-sums in registers while red ones represent group-sums in buffers.}  
    \label{fig:partial_group_sum}
\end{figure}
\subsection{Dataflow in CONV Layers}

COM dataflow in CONV layers varies from existing WS dataflow. Matrix conversion (e.g., im2col) is compulsory in WS dataflow to support convolution operations, which not only requires additional circuits but also greatly increases costs of accessing data in IFMs. We propose a novel dataflow that the matrix conversion is no longer required.

The dimension of a weight tensor in a CONV layer is $K\times K\times C\times M$, where $K$ is the filter size, $C$ is the number of input channels, and $M$ is the number of output channels. In a simple case that $N_c=C$ and $N_m=M$, a slice of a tensor with a size of $C\times M$ is mapped to a CIM array and the complete weight tensor is mapped to $K^2$ tiles. If the size of the CIM array is too small, an array of $\lceil\frac{C}{N_c}\rceil\times\lceil\frac{M}{N_m}\rceil$ tiles are required for a $C\times M$ tensor slice. Therefore, a total number of $K^2\times\lceil\frac{C}{N_c}\rceil\times\lceil\frac{M}{N_m}\rceil$ tiles are allocated for a weight tensor. 

\begin{figure}[hbp]
    \centering  
    \includegraphics[width=0.37\textwidth]{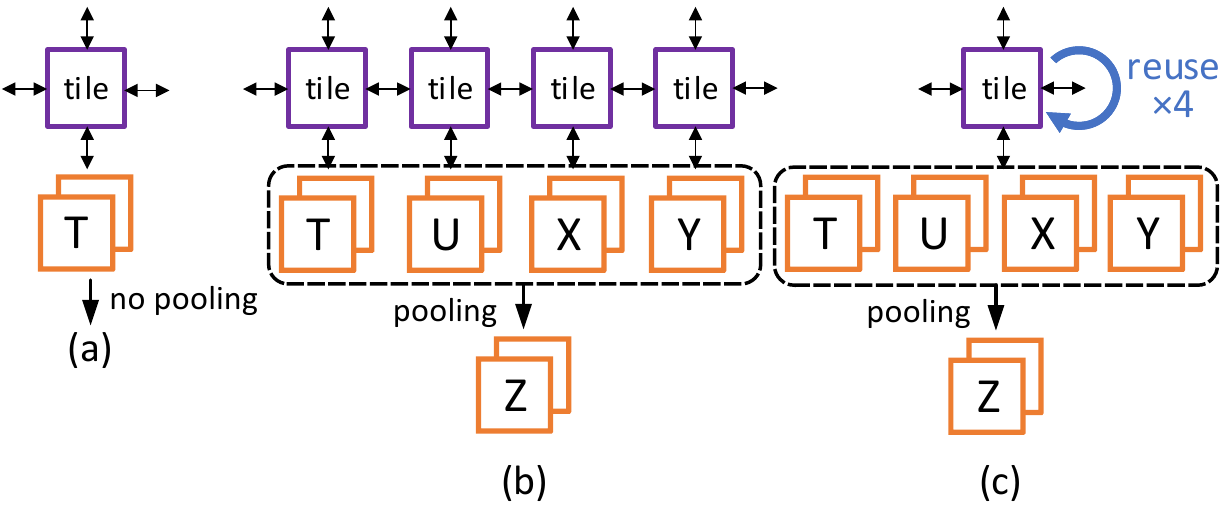}
    \caption{Output in the last tile: weight duplication or block reuse scheme is used to deal with pooling layers.}  
    \label{fig:pooling}
\end{figure}

The computation of a data in an OFM is the sum of point-wise MAC resulting from a sliding window. As shown in \autoref{fig:partial_group_sum} (a), we show the weight mapping strategy. Pixels in kernels are mapped to CIM arrays according to their locations and channels in sequence, which is different from \cite{ISSCC_2021_Jia} that flattens kernels on a single CIM array. \autoref{fig:partial_group_sum} (b) demonstrates the COM dataflow. We define the $N_m$ point-wise MAC result as the partial-sum, $U_1$ to $U_{K^2}$, and row-wise addition result as the group-sum, $U_{g1}$ to $U_{gK}$. Partial-sums and group-sums are sequentially generated and summed up one by one, in different timing and tiles. Partial-sums are generated and transmitted in a pipeline along tiles. Thereby partial- and group-sums are ``computed on the move''. Every $K$ partial-sums ($U_1$ to $U_K$) are summed up as a group-sum ($U_{g1}$). Each group-sum is stored in an ROFM buffer to wait for another group-sum to be ready. In the last tile, $K$ group-sums ($U_{g1}$ to $U_{gK}$) complete accumulation and an activation function is applied in the computation unit of the ROFM.

\begin{table}[b]
    \scriptsize
    \caption{Configuration information summary for Domino under evaluation.}
    \begin{center}
        \begin{tabular}{|c|c|c|c|}
        \hline
            \textbf{Component} & \textbf{Descript.} & \textbf{Energy/Compo.} & \textbf{Area ($\upmu$m$^{2}$) } \\
            \hline
            Buffer & 256B$\times$1  & 281.3 pJ& 826.5 \\
            \hline
            Control circuits & 1 & 10.4 pJ & 1400.6 \\       
            \hline
            \textbf{RIFM total} & \multicolumn{2}{c|}{-}  & 2227.1  \\
            \hline        
            \hline
            Adder & 8b$\times$8$\times$2 & 0.02 pJ/8b & 0.07 \\
            \hline
            Pooling & 8b$\times$8 & 7.7 fJ/8b & 34.06 \\
            \hline
            Activation & 8b$\times$8 & 0.9 fJ/8b & 7.07 \\
            \hline
            Data buffer & 16KiB  &  281.3 pJ & 52896 \\
            \hline
            Schedule table & 16b$\times$128  &  2.2 pJ/16b & 826.5 \\
            \hline
            Input buffer & 64b$\times$2  & 42.1 pJ/64b & 878.9  \\
            \hline
            Output buffer & 64b$\times$2  & 42.1 pJ/64b & 878.9  \\
            \hline
            Control circuits  & 1 &28.5 pJ & 2451.2 \\
            \hline 
            \textbf{ROFM total} &\multicolumn{2}{c|}{-} & 57972.7 \\
            \hline
            \hline
            \textbf{Inter-chip Conn.} & 80Gpbs$\times$8  & 0.55 pJ/b & 8E5 \\
            \hline
        \end{tabular}
        \label{tab:domino_config}
    \end{center}
\end{table}

\begin{table*}[bp]
    \caption{Domino's evaluation results and pairwise comparisons under different DNN models.}
    \begin{center}
        \begin{threeparttable}
            \scriptsize
            \begin{tabular}{|c|c|c|c|c|c|c|c|c|c|c|}
            \hline
                Dataset & \multicolumn{4}{c|}{CIFAR-10}  & \multicolumn{6}{c|}{ImageNet} \\
                \hline
                Model  & \multicolumn{2}{c|}{VGG-11\cite{ISSCC_2021_Jia}} & \multicolumn{2}{c|}{ResNet-18\cite{resnet}} & \multicolumn{2}{c|}{VGG-16\cite{vggnet}} & \multicolumn{4}{c|}{VGG-19\cite{vggnet}} \\
                \hline
                Architecture & \cite{ISSCC_2021_Jia} & Ours & \cite{ISSCC_2020_Yue} & Ours & \cite{ISSCC_2021_Yoon}\textcolor{gray}{\tnote{*1}}  & Ours & \cite{AtomLayer} & Ours & \cite{CASCADE} & Ours \\
                \hline
                \makecell[c]{CIM type} 
                & SRAM  &  SRAM  & SRAM    &  SRAM & ReRAM  &  ReRAM & ReRAM & ReRAM  & ReRAM   & ReRAM \\
                \hline
                \makecell[c]{Technology (nm)} 
                & 16    & 45     & 65      & 45    & 40     &  45      & 32    & 45       & 65      & 45    \\
                \hline
                \makecell[c]{VDD (V)} 
                & 0.8   & 1      & 1       & 1     & 0.9    & 1        & 1     & 1        & 1       & 1     \\
                \hline
                \makecell[c]{Frequency (MHz)} 
                & 200   & 10     & 100     & 10    & 100    & 10       & 1200  & 10       & 1200    & 10  \\
                \hline
                \makecell[c]{Activation \& Weight precision}
                & 4     & 8      & 4       & 8     & 8      & 8  & 16    & 8        & 16      & 8  \\
                \hline
                \makecell[c]{\# of CIM cores/chip \& chips} 
                & 16    &240$\times$5   & 4       & 240$\times$6  & 1      &240$\times$10 & 160  &240$\times$10 & 80 - 112    & 240$\times$10   \\
                \hline
                \makecell[c]{Active area (mm$^2$)} 
                & 17.5    & 343.2   & 5.68   & 655.2  & 0.44      & 381.6     & 6.89  & 192.0    & 0.99    & 125.5      \\
                \hline
                \makecell[c]{Execution time (us)} 
                & 128   & 137.3  & 1890    & 206.3 & 670K  & 3481.8     &  6920  & 3582.9     & n.a.    & 3582.9   \\
                \hline
                \makecell[c]{Power (W)} 
                & 0.15  & 11.03  & 2.78m   & 18.10 & 11.05m  & 4.26     & 4.8 & 8.73      & 3m   & 4.57    \\
                \hline
                \makecell[c]{On-chip data power (W)\textcolor{gray}{\tnote{*2}}} 
                & 0.036   & 3.53(3.50)  & 1.76m   & 2.95(2.93) & 1.47m  & 0.64(0.63)   &  0.54  & 0.72(0.71)    &  0.7m   & 0.72(0.71)   \\
                \hline
                \makecell[c]{Off-chip data power (W)} 
                & 0.06   & 0.34  & n.a.    & 0.10 &  4.76m  & 0.005  & 1.32  & 0.01  & 0.9m    & 0.01   \\
                \hline
                \makecell[c]{CE (TOPS/W)} 
                & 71.39 & 17.22  & 6.91    & 6.30  & 4.15   & 9.29     & 0.68  & 5.73     & 1.96    & 10.95  \\
                \hline
                \makecell[c]{Normalized CE (TOPS/W)\textcolor{gray}{\tnote{*3}}} 
                & 9.53  & 17.22  & 2.82    & 6.30  & 3.92   &9.29     & 2.73  & 5.73     &  6.18   & 10.95   \\
                \hline
                \makecell[c]{Throughput (TOPS/mm$^2$)} 
                & 0.7 & 0.55   & 0.006   & 0.17 & 0.10  & 0.10     & 0.36   & 0.22    & 0.10    & 0.66   \\
                \hline
                \makecell[c]{Normalized throughput (TOPS/mm$^2$)\textcolor{gray}{\tnote{*4}}} 
                & 0.088 & 0.55   & 0.013   & 0.17 & 0.081  & 0.10    & 0.18  & 0.22     & 0.21    & 0.66   \\
                \hline
                \makecell[c]{Images/s/core}
                & 488  & 2604 &  8 & 2604   & n.a.   & 53  &   n.a.  & 53 & n.a.  & 53  \\
                \hline
                \makecell[c]{Accuracy(\%)} 
                & 91.51 & 89.85  & 91.15 &  91.57  & 46     &  70.71   & n.a.  & 72.38    &  n.a.   & 72.38  \\
                \hline
            \end{tabular}
            \begin{tablenotes}
                \item \textcolor{gray}{*1} \textcolor{gray}{Adapted and normalized from average statistics.} 
                \textcolor{gray}{*2} \textcolor{gray}{In parentheses is power of on-chip data movement included.} 
                \textcolor{gray}{*3} \textcolor{gray}{Normalized to 8-bit, 1 V, and 45 nm according to \cite{node-scaling}.}
                \textcolor{gray}{*4} \textcolor{gray}{Normalized to 8-bit, 45 nm.}
            \end{tablenotes}
            \label{tab:domino_performance}
        \end{threeparttable}
    \end{center}
\end{table*}

\subsection{Pooling Dataflow}
Computations of CONV and FC layers are processed within an array of tiles, while computations of pooling layers are performed during data transmission between arrays. If a pooling layer follows a CONV layer, with pooling filter size $K_p = 2$ and pooling stride $S_p = 2$, every four activation results produce a pooling result. As shown in \autoref{fig:pooling} (b), Domino duplicates weights to produce four activation results $T$ to $Y$ in every cycle, which aims to maintain synchronization among layers. When transmitting across tiles, data are compared, and the pooling result $Z$ is produced. \autoref{fig:pooling} (c) shows the block reuse scheme that activation results are computed and stored in the last tile. A comparison is taken when the next activation result is computed. The ROFM outputs a pooling result $Z$ once the comparison in a pooling filter is completed. In this scenario, computation frequency before pooling layers is $4\times$ higher than succeeding blocks. 

\section{Evaluation}
\label{sec:evaluation}
This section evaluates Domino's characteristics and performances in detail. We also compare Domino against other state-of-the-art CIM-based architectures.

\subsection{Experiment Setup}

The configuration of Domino and its tile is displayed in \autoref{tab:domino_config}. Buffer parameters are based on the silicon-proven SRAM array in \cite{sram}. On-chip data transmission energy is simulated by Noxim\cite{Noxim1}, and the rest is analyzed by PrimeTime with a 45 nm CMOS process. The step frequency for the execution of one instruction is 10 MHz and thus the bandwidth between tiles is 40 Gbps. Frequency division multiplexing with 160 MHz clock frequency is implemented in peripheral circuits to reduce the hardware area. Eight 80 Gbps transceivers, adopted from \cite{wireline}, serve as inter-chip connections. The supply voltage is 1 V and the precision is 8 bits. Our model adopts CIM arrays whose size is $256\times256$, and the number of total CIM arrays depends on the scale of neural networks. We run several prevailing CNN models in Domino architecture built by SystemC.

In \autoref{tab:domino_performance}, the counterpart results on the adjacent column are normalized to our aforementioned settings. The array size and bit precision are normalized by linear scaling factors. Let $B_{wt}$, $B_{at}$, $B_{wd}$ and $B_{ad}$ be the weight precision of target architecture, activation precision of target architecture, weight precision of Domino, and activation precision of Domino, respectively. The scaling factor is $\frac{B_{wd} B_{ad}}{B_{wt} B_{at}}$ for MAC and $\frac{B_{ad}}{B_{at}}$ for the rest of operations and data movement. To make a fair energy efficiency comparison, we further normalize technology nodes and supply voltage using equations given in \cite{node-scaling}. In the accuracy simulation, only the quantization error is considered.

\subsection{Performance Results}

\subsubsection{Computational Efficiency}
\label{sec:CE}

Domino achieves higher Computational Efficiency (CE) than its state-of-the-art counterparts. Results show Domino has 77\% (compared to \cite{CASCADE}) to 137\% (compared to \cite{ISSCC_2021_Yoon}) improvement of CE. The reason is Domino largely reduces the energy consumption of both on- and off-chip data movement. Some unique ``skip'' operations appeared in ResNet only affects performances slightly. Data locality and COM dataflow are very efficient in reducing overall energy consumption for CNN inference. Thereby, peripheral energy decreases and system CE increases.

\subsubsection{Throughput}
Domino has an obvious advantage over other architectures in terms of throughput with respect to area. The area per chip is determined by two factors: the area of one tile including substituted CIM arrays and Domino's routers, and the number of tiles according to mapping strategy. We calculate the area of an equivalent CIM array of 256$\times$256 from counterpart models. Throughput is improved by 1.28$\times$ to 13.16$\times$. 

We also compared the inference speed of different neural network models. To make a fair comparison, we normalize the inference speed to one CIM core (images per second per CIM core). Results show that the inference speed is improved by more than five times. This improvement is benefited from layer synchronization, weight stationary, data locality, and COM dataflow, which help to reduce the computing latency and maximize parallelism.

\subsubsection{Power Breakdown}
We break down the total power consumption into three parts: CIM power, on-chip data power and off-chip data power. Because Domino uses others' CIM arrays, power consumption of CIM is not listed. On-chip data power includes on-chip data movement and computation power except CIM, while off-chip data power is responsible for inter-chip communication. When a DNN is too large to be mapped onto a single chip, e.g., ResNet-50, VGGNet, off-chip access is inevitable, involving inter-chip data movement such as IFMs and OFMs. However, as listed in \autoref{tab:domino_performance}, data movement only accounts for a 
small portion (8\% to 32\% for on-chip and 0.1\% to 3\% for off-chip), which means Domino efficiently reduces the overhead of data movement. 

\section{Conclusion}
\label{conclusion}

This paper has presented a customized NoC architecture called Domino with highly localized inter-memory computing for DNNs. Key contributions and innovations can be concluded as follows: (1) Domino changes the conventional NoC tile structure by using dual routers for different usages, and enables substitution of PEs; (2) Domino utilizes an efficient COM dataflow to minimize data movement; and (3) a set of periodical instructions is defined to maximize the data locality. Compared with several conventional architectures, Domino has improved computational efficiency and throughput by 1.77-to-2.37$\times$ and 1.28-to-13.16$\times$, respectively.

\ifCLASSOPTIONcaptionsoff
  \newpage
\fi

\end{document}